\documentclass[reprint,superscriptaddress,amsmath,amssymb,aps,prl,preprintnumbers,longbibliography]{revtex4-2}
\usepackage{latexsym}
\usepackage{comment}
\usepackage[english]{babel}
\usepackage[font={small}]{caption}
\usepackage{epsfig,amssymb,euscript,slashed}
\usepackage[linktocpage]{hyperref}
\usepackage{amsmath}
\usepackage{mathtools}
\usepackage{physics}
\usepackage{extarrows}
\usepackage{xcolor}
\usepackage{float}
\usepackage{graphicx}
\usepackage{caption}
\usepackage{subcaption}
\usepackage{tcolorbox}
\usepackage{MnSymbol}
\usepackage{tikz}
\usetikzlibrary{positioning,shapes.multipart,arrows.meta,decorations.pathreplacing,calligraphy}
\usepackage{adjustbox}
\usepackage{nicematrix}
\usepackage{array,calc}
\usepackage[]{graphicx}
\usepackage{color}
\usepackage{bbm}
\usepackage{fancybox}
\usepackage{ytableau}
\usepackage{eso-pic} 
\usepackage{enumerate}
\usepackage[shortlabels]{enumitem}

\usepackage{pict2e}

\makeatletter
\DeclareRobustCommand{\loplus}{\mathbin{\mathpalette\dog@lsemi{+}}}
\DeclareRobustCommand{\lotimes}{\mathbin{\mathpalette\dog@lsemi{\times}}}
\DeclareRobustCommand{\roplus}{\mathbin{\mathpalette\dog@rsemi{+}}}
\DeclareRobustCommand{\rotimes}{\mathbin{\mathpalette\dog@rsemi{\times}}}

\newcommand{\dog@rsemi}[2]{\dog@semi{#1}{#2}{-90,90}}
\newcommand{\dog@lsemi}[2]{\dog@semi{#1}{#2}{270,90}}
\newcommand{\dog@semi}[3]{%
  \begingroup
  \sbox\z@{$\m@th#1#2$}%
  \setlength{\unitlength}{\dimexpr\ht\z@+\dp\z@\relax}%
  \makebox[\wd\z@]{\raisebox{-\dp\z@}{%
    \begin{picture}(1,1)
    \linethickness{\variable@rule{#1}}
    \roundcap
    \put(0.5,0.5){\makebox(0,0){\raisebox{\dp\z@}{$\m@th#1#2$}}}
    \put(0.5,0.5){\arc[#3]{0.5}}
    \end{picture}%
  }}%
  \endgroup
}
\newcommand{\variable@rule}[1]{%
  \fontdimen8  
  \ifx#1\displaystyle\textfont3\else
    \ifx#1\textstyle\textfont3\else
      \ifx#1\scriptstyle\scriptfont3\else
        \scriptscriptfont3\relax
  \fi\fi\fi
}
\makeatother

\def\Ad{\dot{A}}

\def\St{\tilde{S}}

\def\ty{\tilde{y}}
\def\te{\tilde{\eta}}
\def\tj{\tilde{\jmath}}

\def\ty{\tilde{y}}

\def\lam{\lambda}

\makeatletter
\def\fnum@figure{FIG.~\thefigure}
\def\fnum@table{TABLE~\thetable}
\makeatother

\makeatletter
\g@addto@macro\bfseries{\boldmath}
\makeatother

\def\jt{\tilde{\jmath}}

\def\cA{{\cal A}}

\def\cD{{\cal D}}
\def\cE{{\cal E}}

\def\cH{{\cal H}}

\def\cO{{\cal O}}

\begin{document}

\preprint{\small YITP-25-148}

\title{A New Supersymmetry Index for the D1-D5 CFT}

\author{Marcel R. R. Hughes}
\affiliation{Department of Physics, Nagoya University, Furo-cho, Chikusa-ku, Nagoya 464-8602, Japan}
\author{Masaki Shigemori}
\affiliation{Department of Physics, Nagoya University, 
Furo-cho, Chikusa-ku, Nagoya 464-8602, Japan}
\affiliation{Center for Gravitational Physics,
Yukawa Institute for Theoretical Physics, Kyoto University,
Kitashirakawa Oiwakecho, Sakyo-ku, Kyoto 606-8502, Japan}

\begin{abstract}
We propose a new supersymmetry index for the D1-D5 CFT for $T^4$, relevant to the AdS$_3$/CFT$_2$ correspondence. In a novel formulation of symmetric orbifold CFTs based on the Schur-Weyl duality, we show how this index can be naturally described and its protection is argued based on the detailed nature of exactly marginal operators in these theories. This index is a one-parameter generalization of the standard index and gives more fine-grained information about the structure of microstates than previously available. We demonstrate precise matching of the new index between supergravity and CFT below the black-hole threshold, where the standard index---the modified elliptic genus---is trivial. Above the threshold, we uncover a decomposition of black-hole microstates into distinct sectors, invisible to the modified elliptic genus.
\end{abstract}

\maketitle


\section{1.~~Introduction}
\label{sec:intro}

Symmetric orbifolds of two-dimensional conformal field theories (CFTs) \cite{Dixon:1986jc,Dixon:1986qv} provide an invaluable landscape of theories useful for understanding holography; these universally exhibit many of the properties expected of holographic theories at large $N$,
including the factorization of correlators~\cite{Lunin:2000yv,Pakman:2009zz,Pakman:2009ab}, a Hawking-Page phase transition~\cite{Keller:2011xi,Hartman:2014oaa} and the form of thermal 2-point functions~\cite{Belin:2025nqd}. While sufficient conditions for a CFT to be holographic remain elusive, large classes of symmetric orbifold CFTs were shown to be consistent with a holographic description somewhere in their moduli spaces~\cite{Belin:2019rba,Belin:2020nmp,Benjamin:2022jin}.

A paradigmatic example of symmetric orbifold CFTs in holography is the D1-D5 CFT arising in the AdS$_3$/CFT$_2$ correspondence \cite{Maldacena:1997re}, which has been central to the development of the holographic dictionary.  Below the black-hole threshold energy, agreement between the CFT and the bulk excitation spectra was established~\cite{Maldacena:1998bw, deBoer:1998ip, Deger:1998nm, Larsen:1998xm}, and above the threshold the CFT correctly reproduces the Bekenstein-Hawking entropy of the AdS black hole~\cite{Strominger:1996sh}. Supersymmetry indices \cite{Witten:1982df, deBoer:1998us, Maldacena:1999bp}  are key to such matching, being independent of the coupling relating the symmetric orbifold and gravity regimes.

Recent advances are beginning to reveal the finer structure of the matched spectra: fortuity \cite{Chang:2024zqi, Chang:2025rqy} refines the notion of typicality for black-hole microstates, while CFT technologies \cite{Gava:2002xb, Guo:2019ady, Guo:2020gxm, Hampton:2018ygz, Guo:2019pzk, Benjamin:2021zkn, Guo:2022ifr, Gaberdiel:2024nge} allow one to determine whether given microstates remain supersymmetric when interactions are turned on. Existing supersymmetry indices are of limited use here: defined purely from the supersymmetry algebra, they are largely insensitive to the details of interactions.

In this Letter, we introduce a new Schur-Weyl-based formulation for symmetric orbifold CFTs which decomposes the BPS Hilbert space into sectors of definite symmetry with respect to the action of the symmetric group. For the D1-D5 CFT on $T^4$ we demonstrate that this formulation leads to a new formula for the standard supersymmetry index of this theory---the modified elliptic genus (MEG)--- which is more transparent with regards to the contributing states. In this framework we are then motivated to propose a new supersymmetry index---the \emph{resolved elliptic genus (REG)}---that incorporates selection rules based on the nature of the interaction in symmetric orbifold CFTs. For the D1-D5 CFT on $T^4$ we demonstrate that the REG provides much more information about the structure of microstates than the MEG\@.

\section{2.~~Symmetric orbifold CFTs}

An $M^N/S_N$ symmetric orbifold CFT with central charge $c=Nc_0$ can be obtained by starting with the ``seed'' CFT $M$ with central charge $c_0$, taking its $N$-fold tensor product, and viewing each copy (or “strand”) as a separate worldsheet carrying the fields of $M$.
Orbifolding by $S_N$ then allows permutations of these copies as we go around the strand, effectively gluing some of the strands together into a longer strand.  Requirement of $S_N$ invariance leads \cite{Dixon:1986jc} to the Hilbert space decomposing into ``twist sectors'' $\mathcal{H}(M^N/S_N)\cong\bigoplus_{[g]}\mathcal{H}_{[g]}$ labeled by conjugacy classes $[g]$. For $S_N$ these conjugacy classes are labeled by partitions $\{n_k\}$ of $N$, satisfying
 $\sum_{k=1}^N k n_k=N$,
where $n_k$ is the number of strands of length $k$. 
$\mathcal{H}_{[g]}$ has the interpretation \cite{Dijkgraaf:1996xw} as a multi-string Hilbert space where a strand of length $k$ represents a string wrapping $k$ times around the $S^1$ on which the CFT lives.  It is then natural to introduce a ``covering'' Hilbert space, or Fock space, $\cH$, in which we include all values of $N$ by summing over $n_k\in\mathbb{Z}_{\ge 0}$.  The original Hilbert space at fixed $N$ is obtained by projection onto the subspace with total strand number $N$.

Besides operators coming from the seed theory, symmetric orbifold CFTs contain twist operators $\sigma_k$, labeled by cyclic permutations of order $k$, which generate twist-sector ground states. In particular, $\sigma_2$, which splits a strand into two shorter strands or glues together two strands into a longer one, is used to construct a marginal deformation operator and plays an essential role in linking symmetric orbifold CFTs to a holographic setup.

In the context of holography, the most studied example of symmetric orbifold CFTs is the D1-D5 CFT \cite{Aharony:1998qu,David:2002wn, Avery:2010qw} where the seed theory is an $\mathcal{N}=(4,4)$ supersymmetric sigma model on $M=T^4$ or $K3$. Here we will focus on the $T^4$ theory, for which the bosonic symmetries include the left- and right-moving R-symmetries $SU(2)_L\times SU(2)_R$ and an $SU(2)_1\times SU(2)_2$ symmetry broken \footnote{The $SU(2)_1\times SU(2)_2$ symmetry coming from the isometries of the $T^4$ target space are broken by states carrying non-trivial momentum and winding on this torus. We will be solely focusing on the BPS states of the CFT which are contained in the zero momentum and winding sector and so use these $SU(2)$ to organize our states.} by the compactness of $T^4$. Per strand, the field content consists of four free bosons and four free left- and right-moving fermions. The RR (Ramond-Ramond) ground states on a strand of length $k$ are tensor products of the left-moving ground states $\ket*{\alpha}_k,\ket*{\Ad}_k$ and the right-moving ground states $\ket*{\dot\alpha}_k, \ket*{\Ad}_k$, where $\alpha,\dot{\alpha},\Ad$ are doublet indices for $SU(2)_L,SU(2)_R,SU(2)_2$, respectively. The states $\ket*{\alpha}_k,\ket*{\dot\alpha}_k$ are bosonic while the $\ket*{\Ad}_k$ are fermionic in our conventions.
Each of these ground states can be excited by oscillator modes of the free bosons and fermions. A state in the full theory is then obtained by tensoring states for all strands together and symmetrizing with respect to the strand label.
In this Letter, we will focus on 1/4-BPS states whose right-moving sector is in a R ground state and the left-moving sector can be in an arbitrary state.

While the above is the standard description of the Ramond ground states of this theory and the representations in which they transform, for our purposes it will prove important to distinguish the action of $SU(2)_2$ on the left-moving and right-moving parts of states separately. We define the group $\widetilde{SU}(2)_2$ as the right-moving action of $SU(2)_2$, under which the right-moving $\ket*{\Ad}$ Ramond ground states transform as a doublet and the analogous left-moving ground states as singlets.

\section{3.~~A Schur-Weyl formulation}

An alternative construction  \footnote{For the mathematical background for this section, such as the Schur-Weyl duality, Schur functions, and Cauchy identities, see \cite{Fulton:1991} for the purely bosonic case and \cite{469926,BERELE1987118,BERELE1985225} for supersymmetric extensions.} of the covering Hilbert space $\cH$, and one that will prove useful to construction of the new, more refined, index with which this Letter is concerned, first factors the theory into left- and right-moving sectors \footnote{For other approaches to symmetric orbifold CFTs based on left-right factorization and the Schur-Weyl duality, see {\it e.g.}~\cite{Jevicki:2015irq}.}. Let $V_k$ be the left-moving Hilbert space on a length-$k$ strand, for which the covering space is $V = \bigoplus_{k=1}^\infty V_k$. Containing infinitely many bosonic and fermionic states, $V$ is naturally a fundamental representation space of $GL(\infty|\infty)$.
For the right-movers, we take a $k$-independent single-strand Hilbert space of R ground states, $\tilde{V}$: right-moving BPS Hilbert spaces for different strand lengths are assumed to be all isomorphic. If $\tilde{V}$ contains $b$ bosonic and $f$ fermionic states, it is naturally a fundamental representation space of $GL(b\,|\,f)$.

The $n$-strand Hilbert spaces for the left- and right-movers, by the Schur-Weyl duality, decomposes as
\begin{equation}
    V^{\otimes n}\cong
    \bigoplus_{\lambda\,\vdash n}\, V_\lambda \otimes M_\lambda\ 
    ,\quad\ 
    \tilde{V}^{\otimes n}\cong
    \bigoplus_{\tilde{\lambda}\,\vdash n}\, \tilde{V}_{\tilde{\lambda}} \otimes M_{\tilde{\lambda}}\ ,
\label{eq:V^n}
\end{equation}
where the sums are over Young diagrams $\lambda,\tilde{\lambda}$, each with $n$ boxes (as denoted by $\lambda,\tilde{\lambda}\vdash n$). In \eqref{eq:V^n}, $V_\lambda\subset V^{\otimes n}$ and $\tilde{V}_{\tilde{\lambda}}\subset \tilde{V}^{\otimes n}$ are irreducible $GL(\infty|\infty)$- and $GL(b\,|\,f)$-modules with representations labeled by $\lambda$ and $\tilde{\lambda}$, obtained by the action of the so-called Schur functor. More physically, if $V$ is the space of vectors $T^I$ where $I=1,\dots,\dim V$, then $V_{\lambda}$ is the subspace of tensors $T^{I_1\dots I_n}$ with symmetry dictated by $\lambda$ \footnote{Note that swapping fermionic indices yields an extra minus sign} (and similarly for $\tilde{V}_{\tilde{\lambda}}$). $M_\lambda$ and $M_{\tilde{\lambda}}$ are irreducible $S_n$-representations called Specht modules, labeled by $\lambda$ and $\tilde{\lambda}$, respectively.

Let us assume that the physical $n$-strand Hilbert space $\cH_n$ is the $S_n$-invariant subspace of $V^{\otimes n}\otimes \tilde{V}^{\otimes n}$ (we will comment on this assumption later).  Since the $S_n$ product representation $M_{\lambda}\otimes M_{\tilde{\lambda}}$ contains the trivial representation if and only if $\lambda=\tilde{\lambda}$, we find
\footnote{This simple form is because $\tilde{V}$ does not involve summation over the strand number. If it did, we would also have to project \eqref{eq:Hn_LxR} onto the subspace in which the left and right strand numbers are the same.}
\begin{equation}
    \cH_n=\bigoplus_{\lambda\,\vdash n} V_\lambda \otimes \tilde{V}_{\lambda}\ ,
\label{eq:Hn_LxR}
\end{equation}
where we dropped trivial Specht modules.
Then the covering Hilbert space, summed over all strand numbers, is 
\begin{equation}
    \cH = \bigoplus_{\lambda} V_\lambda \otimes \tilde{V}_{\lambda}\ ,
\label{eq:H_LxR}
\end{equation}
where the sum is over $\lambda$ with arbitrary numbers of boxes. The Hilbert space for fixed $N$ is obtained by projection onto the subspace with total strand number $N$.

Let the (super)character of $V$---or physically the left-moving single-strand (signed) partition function---be
\begin{equation} \label{eq:1-strand_char_L}
    z(x|x')=\sum_i x_i - \sum_{i'} x'_{i'}\ ,
\end{equation}
where $i$ and $i'$ run over bosonic and fermionic states respectively in $V$, and $x_i$ and $x'_{i'}$ are the eigenvalues of an arbitrary operator $g\in GL(\infty|\infty)$ in principle, but for applications to indices, $g=p^{\hat{k}}q^{L_0-c/24}y^{2J^3_0}$ where $\hat{k}$ is the strand length operator, $L_0$ is a Virasoro generator, and $J^3_0$ is the Cartan generator of $SU(2)_L$. The minus sign for the second term corresponds to having $(-1)^F$, with $F$ being the fermion number operator, in the character's trace. Analogously, the character of $\tilde{V}$ (or the right-moving single-strand partition function), assumed to be $k$-independent, is
\begin{equation} \label{eq:1-strand_char_R}
   \tilde{z}(\tilde{x}|\tilde{x}')
    =
        \sum_{\tilde{\imath}} \tilde{x}_{\tilde{\imath}}
   -     \sum_{\tilde{\imath}'} \tilde{x}'_{\tilde{\imath}'}\ ,
\end{equation}
where the chosen operator is typically $\tilde{g} = \tilde{y}^{2\tilde{J}^3_0}\in GL(b\,|\,f)$ with $\tilde{J}^3_0$ being the Cartan generator of $SU(2)_R$.
Then the character---or physically the (signed) partition function---of the covering Hilbert space \eqref{eq:H_LxR} is
\begin{equation}
    \mathcal{Z}=\sum_{\lambda}
        S_\lambda(x|x')\,
    \St_{\lambda}(\tilde{x}|\tilde{x}')\ ,
        \label{eq:Z=SS}
\end{equation}
where $S_\lambda(x|x')$, $\St_\lambda(\tilde{x}|\tilde{x}')$ are the characters of $V_\lambda$, $\tilde{V}_\lambda$, and are given by super Schur functions; see Supplemental Material.

\section{4.~~The new index}

Before applying the formalism of the previous section, we first review the standard index for the D1-D5 CFT\@. A supersymmetry index is invariant under continuous changes of the coupling; it is defined so that states that can recombine---and ``lift''---contribute zero \cite{Witten:1982df}. For the D1-D5 CFT on $T^4$, the appropriate index is the MEG \cite{Maldacena:1999bp}, defined by the R-R sector trace over the free-theory BPS Hilbert space
\begin{equation}
\cE_N(q,y)=
    \cD\tr[(-1)^Fq^{L_0-c/24}y^{2J^3_0}\tilde{y}^{2\tilde{J}^3_0}]\ ,
    \label{eq:MEG_T^4}
\end{equation}
where $(-1)^F = (-1)^{2(J^3_0-\tilde{J}^3_0)}$ provides a $\mathbb{Z}_2$ grading by spin statistics and the differential operator acts as ${\mathcal D} \coloneq \frac12 (\tilde{y}\partial_{\tilde y})^2$ followed by setting all right-moving fugacities to one: $\ty=\te=1$. Although the 1/4-BPS spectrum of the D1-D5 CFT is not invariant under turning on couplings of the 20 exactly marginal operators of this theory, the MEG is protected: quartets of free-theory short multiplets that combine into a long multiplet in the interacting theory contribute zero.

To compute this MEG, one first writes the single-strand left-moving character \eqref{eq:1-strand_char_L} as 
\begin{align}
    z(p,q,y)=  \sum_{k,m,l}c(k,m,l)\,p^k q^m y^l\ ,
    \label{eq:1-strand_z(p,q,y)}
\end{align}
where $k\ge 1$, $m\ge 0$, $l\in\mathbb{Z}$. This is an expanded form of the condensed expression \eqref{eq:1-strand_char_L}, where the eigenvalues $x_i,x'_{i'}$ correspond to $p^k q^m y^l$. The (signed) degeneracies satisfy $c(k,m,l)=c(km,l)$ \cite{Dijkgraaf:1996xw}, with $c(m,l)$ defined by the seed theory left-moving partition function
\begin{equation}
    \sum_{m,l}c(m,l)\,q^m y^l = -\left(\frac{\vartheta_1(\nu,\tau)}{\eta(\tau)^3}\right)^{2} \ .
    \label{eq:1-strand_z(q,y)}
\end{equation}
where $y=e^{2\pi i\nu}$, $q=e^{2\pi i\tau}$ and the definitions for the Jacobi theta and Dedekind eta functions are given in \eqref{eq.theta1etaDef}.
Then the generating function for $\cE_N$ is given by \cite{Maldacena:1999bp}
\begin{align}
    \cE(p,q,y)\coloneq 
    \!\sum_{N=0}^\infty p^N\cE_N(q,y)
    =\!\sum_{k,m,l} \!\!\frac{c(k,m,l)\,p^k q^m y^l}{(1-p^k q^m y^l)^2} .
    \label{eq:MEG_MMS}
\end{align}

While this is the typical presentation of the MEG, we can now instead work with the Schur-Weyl formalism of the previous section. In the D1-D5 CFT for $T^4$, the left-moving covering Hilbert space is given by $V={\rm span}\{\cO \ket*{\alpha}_k, \cO\ket*{\Ad}_k\}_{k,\cO}$ where $\cO$ represents arbitrary left-moving bosonic and fermionic excitations. The right-moving Hilbert space is taken to be the $k$-independent space $\tilde{V}={\rm span}\{\ket*{\dot{\alpha}},\ket*{\Ad}\}$. Having two bosonic and two fermionic states, $\tilde{V}$ is naturally a $GL(2|2)$ representation space and $\tilde{V}_{\lambda}$ vanishes unless $\lambda_3\le 2$, with $\lambda_i$ being the $i$th row length of the Young diagram $\lambda$. In other words, $\lam$ is constrained to be of the $(2|2)$ hook type which we denote by $\lam\in H(2|2)$ \footnote{Generally, if a vector space $V$ contains $b$ bosonic and $f$ fermionic states, $V_\lambda$ vanishes unless $\lambda_{b+1}\le f$ (the $(b|f)$-hook condition) and so $\lam\in H(b|f)$. This is because we cannot anti-symmetrize more than $b$ bosonic indices and symmetrize more than $f$ fermionic indices in the tensor $T^{I_1\dots I_n}$.}. Noting that the trace in \eqref{eq:MEG_T^4} is nothing but the (signed) partition function for the theory at fixed $N$, the generating function $\mathcal{E}(p,q,y)$ can be obtained by the action of $\mathcal{D}$ on \eqref{eq:Z=SS}, the partition function of the covering Hilbert space. We then find a Schur-Weyl expression for the MEG generating function:
\begin{equation}
    \cE(p,q,y)=\sum_{\lambda\in H(2|2)} S_\lambda(p,q,y) \, \cD \St_\lambda\ ,
    \label{eq:E=sum_SDS}
\end{equation}
where here $S_\lambda(p,q,y)$ is a Schur function for the single-strand character \eqref{eq:1-strand_z(p,q,y)} and $\St_\lambda(\ty,\te)$ is a Schur function for the $GL(2|2)$ character
\begin{equation} \label{eq.gl22Char}
    \tilde{z}(\ty,\te) = \tilde{y} + \tilde{y}^{-1} - \te - \te^{-1} \ .
\end{equation}
In writing this we have kept only the fugacities $\ty$ and $\te$ respectively for the $SU(2)_R$ and $\widetilde{SU}(2)_2$ subgroups of this $GL(2|2)$, which in the language of \eqref{eq:1-strand_char_R} is a supertrace of the operator $\tilde{g}= \ty^{2\tilde{J}^3_0}\te^{2\check{J}^3_0}$, where $\check{J}^3_0$ is a generator of $\widetilde{SU}(2)_2$. While the fugacity $\te$ is not necessary for defining the MEG (and  we set $\te=1$ when working with the MEG as otherwise it would not be protected), we include it in the right-moving character for later convenience. 

While $\St_\lambda(\ty,\te)$ is non-vanishing only if $\lambda_3\le 2$, $\cD \St_\lambda$ is non-vanishing only if $\lambda_2\le 1$, namely if $\lambda\in H(1|1)$ is a single-hook Young diagram (Fig.~\ref{fig:hook_diag}), and its value in that case is
\begin{equation} \label{eq:GL22MEGcontribution}
    \cD \St_\lambda = (-1)^{\rho_\lambda-1}n_\lambda\ ,
\end{equation}
where $\rho_\lambda$ is the number of rows and $n_\lambda$ is the number of boxes in $\lambda$.
Therefore, we arrive at an alternative expression for the MEG:
\begin{equation}
   \cE(p,q,y)=\sum_{\lambda\in H(1|1)} S_\lambda(p,q,y) 
    \,\, (-1)^{\rho_\lambda-1} n_\lambda\ .
    \label{eq:MEG_as_hook_sum}
\end{equation}

\begin{figure}[tb]
\begin{tikzpicture}[scale=0.75]
\draw (0,0) rectangle +(0.2,-0.2);
\draw (0.2,0) rectangle +(0.2,-0.2);
\draw (0.4,0) rectangle +(0.2,-0.2);
\draw (0.6,0) rectangle +(0.2,-0.2);
\draw (0.8,0) rectangle +(0.2,-0.2);
\draw (1.0,0) rectangle +(0.2,-0.2);
\draw (1.2,0) rectangle +(0.2,-0.2);
\draw (0,-0.2) rectangle +(0.2,-0.2);
\draw (0,-0.4) rectangle +(0.2,-0.2);
\draw (0,-0.6) rectangle +(0.2,-0.2);
\draw[decorate,decoration={brace,mirror,amplitude=6pt}]
       (-0.05,0) -- (-0.05,-0.8)
       node[midway, left=5pt]{$\rho_\lambda$};
\draw[decorate,decoration={brace,amplitude=6pt}]
       (0,0.05) -- (1.4,0.05)
       node[midway, above=5pt]{$n_\lambda-\rho_\lambda+1$};
\end{tikzpicture}
\caption{\raggedright\sl A single-hook Young diagram $\lambda\in H(1|1)$ with $n_\lambda$ boxes and $\rho_\lambda$ rows.
\label{fig:hook_diag}}
\end{figure}

While this new form of the generating function of the MEG is interesting in its own right, it also invites us to consider how the index can be refined by considering superselection sectors preserved by the deformed right-moving supercharge. The operator responsible for generating a long supersymmetry multiplet of the first-order deformed theory out of four short multiplets of the undeformed theory is referred to as the Gava-Narain operator $\mathcal{G}^{\dot{+}A}_{0}$ (and its hermitian conjugate $\sim\mathcal{G}^{\dot{-}A}_{0}$) \cite{Gava:2002xb,Guo:2019pzk}. As denoted by its various indices, the Gava-Narain operator carries right-moving R-charge of $+\frac12$ and transforms as a doublet of the $SU(2)_1$. Related to the first-order deformed right-moving supercharge, the Gava-Narain operator involves the twist operator $\sigma_2$ and therefore does not preserve the usual twist-sector decomposition of the free orbifold theory. It does, however, preserve the left-moving contracted large $\mathcal{N}=4$ symmetry algebra, as well as the Clifford algebra $cl_{4}$ generated by the zero modes (in the Ramond sector) of the right-moving free fermions acting diagonally on the strand structure~\cite{Guo:2019pzk,Guo:2020gxm}. This Clifford algebra is a subalgebra of $gl(2|2)$, the algebra of the group $GL(2|2)$ acting on the space of right-moving Ramond ground states. The right-moving action of the Gava-Narain operator therefore maps between entire representations of $cl_{4}$.
Noting also that the Gava-Narain operator does not carry any charge under the $\widetilde{SU}(2)_2$, one is motivated to organize states into representations of the subalgebra $\mathcal{A} \equiv (su(2)_R\oplus \widetilde{su}(2)_2) \loplus cl_{4}$ rather than of the full $gl(2|2)$ (here $h\loplus g$ means a semidirect sum with $g$ being the ideal).  Irreducible representations of the algebra $\mathcal{A}$ are labeled using the spin $\tj\in\mathbb{Z}_{\ge 0}/2$ under $su(2)_R$ and $\tj_2\in\mathbb{Z}_{\ge 0}/2$ under $\widetilde{su}(2)_2$, with the character of one such representation being
\begin{equation} \label{eq.AalgChars}
    \tilde{\chi}^{\mathcal{A}}_{\tj,\tj_2}(\ty,\te) = (-1)^{2\tj_2}\big[ \tilde{\chi}_{\frac12}(\ty) - \tilde{\chi}_{\frac12}(\te)\big] \tilde{\chi}_{\tj}(\ty)\tilde{\chi}_{\tj_2}(\te) \ ,
\end{equation}
where $\tilde{\chi}_{j}(X)=X^{2j}+\cdots+X^{-2j}$ is a spin $j$ $su(2)$ character. We define $\tilde{\chi}^{\mathcal{A}}_{\tj,\tj_2}$ to be vanishing if $\tj<0$ or $\tj_2<0$.

We are then motivated to decompose the partition function \eqref{eq:Z=SS} into the characters \eqref{eq.AalgChars} and define sectors based upon the $\tj_2$ charge. Note that this $\widetilde{su}(2)_2$ representation is defined after factoring out the Clifford quartet structure, within which states have different $\widetilde{su}(2)_2$ (and $su(2)_R$) charges \footnote{Namely, this is \emph{not} the same as refining the MEG by naively inserting a fugacity for $\widetilde{su}(2)_2$ in the index's trace, which would \emph{not} lead to a protected index.}. To do this it is convenient to parametrize the Young diagram $\lam\in H(2|2)$ labeling the representation of $gl(2|2)$ using Frobenius coordinates: \textit{i.e.}, $\lam = (a_1|b_1)$ for $d_\lambda=1$ and 
$\lam = (a_1,a_2|b_1,b_2)$ for $d_\lambda=2$,
where $d_{\lam}$ is the number of hooks in the diagram $\lam$, or in other words the largest integer $s$ for which $\lam_s\geq s$ (see Fig.~\ref{fig:frobenius_coords}).
%
\begin{figure}
    \centering
\begin{tabular}{c@{\hspace{5ex}}c}
\begin{tikzpicture}[scale=1.3]
 \draw (0  ,0) rectangle +(1ex,1ex);
 \draw (1ex,0) rectangle +(1ex,1ex);
 \draw (2ex,0) rectangle +(1ex,1ex);
 \draw (3ex,0) rectangle +(1ex,1ex);
 \draw (4ex,0) rectangle +(1ex,1ex);
 \draw (0,-1ex) rectangle +(1ex,1ex);
 \draw (0,-2ex) rectangle +(1ex,1ex);
 \draw (0,-3ex) rectangle +(1ex,1ex);
\draw[latex-latex] (1ex,0.5ex) -- +(4ex,0) node [right,xshift=-0.3ex] {$a_1$};
\draw[latex-latex] (0.5ex,0) -- +(0,-3ex) node [below,yshift=0.3ex] {$b_1$};
\end{tikzpicture}
&
\begin{tikzpicture}[scale=1.5]
 \draw (0  ,0) rectangle +(1ex,1ex);
 \draw (1ex,0) rectangle +(1ex,1ex);
 \draw (2ex,0) rectangle +(1ex,1ex);
 \draw (3ex,0) rectangle +(1ex,1ex);
 \draw (4ex,0) rectangle +(1ex,1ex);
 \draw (5ex,0) rectangle +(1ex,1ex);
 \draw (0  ,-1ex) rectangle +(1ex,1ex);
 \draw (1ex,-1ex) rectangle +(1ex,1ex);
 \draw (2ex,-1ex) rectangle +(1ex,1ex);
 \draw (3ex,-1ex) rectangle +(1ex,1ex);
 \draw (4ex,-1ex) rectangle +(1ex,1ex);
 \draw (0,-2ex) rectangle +(1ex,1ex);
 \draw (1ex,-2ex) rectangle +(1ex,1ex);
 \draw (0,-3ex) rectangle +(1ex,1ex);
 \draw (1ex,-3ex) rectangle +(1ex,1ex);
 \draw (0,-4ex) rectangle +(1ex,1ex);
\draw[latex-latex] (1ex,0.5ex) -- +(5ex,0) node [right,xshift=-0.3ex] {$a_1$};
\draw[latex-latex] (2ex,-0.5ex) -- +(3ex,0) node [right,xshift=-0.3ex,yshift=-0.5ex] {$a_2$};
\draw[latex-latex] (0.5ex,0) -- +(0,-4ex) node [below,yshift=0.3ex] {$b_1$};
\draw[latex-latex] (1.5ex,-1ex) -- +(0,-2ex) node [below,yshift=0.3ex,xshift=0.7ex] {$b_2$};
\end{tikzpicture}
\\
(a)&(b)
\end{tabular}
    \caption{Frobenius coordinates for Young diagrams in the case of (a) diagrams with $d_\lambda=1$, $\lambda=(a_1|a_2)\in H(1|1)$ and (b) diagrams with $d_\lambda=2$, $\lambda=(a_1,a_2|b_1,b_2)\in H(2|2)$ with $a_1>a_2$ and $b_1>b_2$.}
    \label{fig:frobenius_coords}
\end{figure}
The decomposition of a $gl(2|2)$-character $S_{\lam}$ into $\cA$-characters can be shown to be given by
\begin{align} \label{eq.gl22toADef}
    \tilde{S}_{\lam} = \begin{cases}
        \ \tilde{\chi}^{\mathcal{A}}_{\frac{a_1}{2},\frac{b_1}{2}} + \tilde{\chi}^{\mathcal{A}}_{\frac{a_1-1}{2},\frac{b_1-1}{2}} & \text{if } d_{\lam} = 1\\[3ex]
        \ \tilde{\chi}^{\mathcal{A}}_{\frac{a_{12}-1}{2},\frac{b_{12}-2}{2}} + \tilde{\chi}^{\mathcal{A}}_{\frac{a_{12}-2}{2},\frac{b_{12}-1}{2}} \\
        \qquad + \tilde{\chi}^{\mathcal{A}}_{\frac{a_{12}}{2},\frac{b_{12}-1}{2}} + \tilde{\chi}^{\mathcal{A}}_{\frac{a_{12}-1}{2},\frac{b_{12}}{2}} & \text{if } d_{\lam} = 2
    \end{cases} \ \,,
\end{align}
where $a_{12}=a_1-a_2$ and $b_{12}=b_1-b_2$.
Therefore, the $\lambda$-sector decomposition
of the MEG \eqref{eq:E=sum_SDS} further breaks down into $\cA$-sectors as
\begin{align}
    \cE(p,q,y)=
\sum_{\lambda\in H(2|2)}
\sum_{(\tj,\tj_2)\subset\lambda} S_\lambda(p,q,y)\,\cD \tilde{\chi}^{\cA}_{\tj,\tj_2}
\label{MEGdecompA}
\end{align}
where $\sum_{(\tj,\tj_2)\subset\lambda}$ means summation over $\cA$-representations contained in the $gl(2|2)$ representation labeled by $\lambda$ according to \eqref{eq.gl22toADef}.
Here
\begin{equation} \label{eq.AcharMEG}
    \mathcal{D} \tilde{\chi}^{\mathcal{A}}_{\tj,\tj_2}
    = (-1)^{2\tj_2}(2\tj+1)(2\tj_2+1) \ .
\end{equation}
As stated above, the Gava-Narain operator generates long multiplets of the deformed theory out of free-theory short multiplets in isomorphic $\widetilde{su}(2)_2$ representations. Therefore, we can define a ``resolved'' version of the MEG 
by summing only over terms in
\eqref{MEGdecompA} that have a particular value of $\tj_2$:
\begin{align} \label{eq.REG1}
    &\mathcal{E}_{\tj_2}(p,q,y) \coloneq \!\!\sum_{\substack{\lambda\in H(2|2)\\\tj_2\subset\lam}}\sum_{\tj_{\lam}} S_\lambda(p,q,y) \cD \tilde{\chi}^{\cA}_{\tj_{\lam},\tj_2}(\ty,\te) \notag\\
    &= \!\!\sum_{\substack{\lambda\in H(2|2)\\\tj_2\subset\lam}}\!\! S_\lambda(p,q,y) (-1)^{\tj_2}(2\tj_2+1) \sum_{\tj_{\lam}} (2\tj_{\lam}+1) \ ,
\end{align}
where the condition ``$\tj_2\subset \lambda$'' means that the $gl(2|2)$-representation labeled by $\lambda$ must contain $\cA$-representations with $\widetilde{su}(2)_2$ charge $\tj_2$, and $\tj_\lambda$ runs over values of the $su(2)_R$ label $\tj$ that appear in the $gl(2|2)$ representation $\lambda$.
%
%
We refer to this as the \emph{resolved elliptic genus (REG)}\@.
Explicitly, the first several terms in \eqref{eq.REG1} are
\begin{align}
\ytableausetup{boxsize=0.5ex}
\cE_{\tj_2=0}&=S_{\ydiagram{1}}
+2S_{\ydiagram{2}}
+3S_{\ydiagram{3}}
+S_{\ydiagram{2,1}}
+4S_{\ydiagram{4}}
+2S_{\ydiagram{3,1}}
+2S_{\ydiagram{2,2}}
+\cdots,\notag\\
\cE_{\tj_2=\frac12}
&=-2S_{\ydiagram{1,1}}
-4S_{\ydiagram{2,1}}
-6S_{\ydiagram{3,1}}
-2S_{\ydiagram{2,2}}
-2S_{\ydiagram{2,1,1}}
+\cdots,\\
\cE_{\tj_2=1}&=
3S_{\ydiagram{1,1,1}}
+6S_{\ydiagram{2,1,1}}
+\cdots,\quad
\cE_{\tj_2=\frac32}=
-4S_{\ydiagram{1,1,1,1}}+\cdots,
\quad \dots\notag
\end{align}
where we showed only terms with up to four boxes.
Since $\cE_{\tj_2}(p,q,y)$ is a generating function, the REG for fixed $N$, which we denote by $\cE_{N,\tj_2}(q,y)$ (where $\tj_2 = 0,\frac12,\dots,\frac{N-1}{2}$), can be found by expanding $\cE_{\tj_2}(p,q,y)$ in $p$.

Another form for this generating function of REG---in the style of \eqref{eq:MEG_MMS}---can also be found by a Cauchy identity for the super Schur functions~\eqref{eq:Cauchy_id_general_app}. One can derive the closed-form expression 
 \begin{align}
    &\cE(p,q,y,x) 
    \coloneq\sum_{\jt_2}\mathcal{E}_{\tj_2}(p,q,y)\frac{x^{2\tj_2+1}+x^{-2\tj_2-1}}{2}
    \notag\\
    &= \big(x^{\frac12} - x^{-\frac12}\big)^{-2} \notag\\
    & \times \prod_{k,m,l} \left(\frac{(1-p^k q^m y^l x)(1-p^k q^m y^l x^{-1})}{(1-p^k q^m y^l)^2}\right)^{c(k,m,l)} 
    \label{eq:REG_DMVV}\\    
    &\times\left[1 + \frac{(x-x^{-1})^{2}}{2} \sum_{k,m,l} 
    \frac{c(k,m,l)\, p^k q^m y^l}{(1- p^k q^m y^l x)(1- p^k q^m y^l x^{-1})}\right] \ ,
\notag
\end{align}
where the $c(k,m,l)$ are defined in \eqref{eq:1-strand_z(p,q,y)}. 
Upon setting $x=1$ this reduces to the MEG \eqref{eq:MEG_MMS} (up to some irrelevant divergent terms at order $p^0$).

Since the arguments made in this section about lifting are made with reference to the first order deformation of the symmetric orbifold theory by one of the twisted-sector moduli, the first order BPS spectrum needs to be exact for the REG to be a fully protected index. While this has long been believed to be true in a wide number of theories, recent work~\cite{Chang:2025mqp} casts some doubt on this for $\mathcal{N}=4$ super Yang-Mills. We also note that our use of representations of the $\widetilde{su}(2)_2\subset\mathcal{A}$ algebra to define superselection sectors for the first order deformed supercharges and label REGs with them is natural only from the free orbifold point. Away from this point the $\widetilde{su}(2)_2$ charge is not a good quantum number and so it is likely that the REG do not provide a meaningful separation of states at a generic point in moduli space. Nonetheless, the REG does provide a meaningful division of the free orbifold spectrum in a manner consistent with first-order lifting and which allows for detailed matching with the KK spectrum in the bulk, as we now demonstrate.

\section{5.~~Application to ${\rm AdS}_3$/CFT$_2$}

Given the new REG defined above for the symmetric orbifold CFT of $T^4$, one can also apply it to the space of states dual to supergravitons in AdS$_3\times S^3$ (or for large enough conformal dimensions, to superstrata backgrounds \cite{Shigemori:2020yuo}). Supergraviton states are naturally defined in the Neveu-Schwarz (NS) sector of the CFT and the left-moving single-strand supergraviton Hilbert space is a restriction of the full NS Hilbert space, $V_{\mathrm{SG}}=\mathrm{span}\{\mathcal{O}_g\ket{\alpha}_k,\mathcal{O}_g\ket*{\dot{A}}_k\}_{k,\mathcal{O}_g}$, where the $\mathcal{O}_g$ are excitations using modes only of the $SU(1,1\,|\,2)$ global (anomaly-free) subalgebra of the full $\mathcal{N}=4$ symmetry algebra \cite{Maldacena:1999bp}. The right-moving single-strand Hilbert space $\tilde{V}$ is the same as the CFT’s Ramond ground-state space
\footnote{The mixed NS-R sector that we consider here allows both an easy comparison to the supergraviton spectrum as well as preserving the simplicity of the right-moving characters in the R sector that we took advantage of in Sec.~3.}. The REG generating function for supergraviton states is again given by \eqref{eq.REG1} (or by \eqref{eq:REG_DMVV}), but now $S_{\lambda}(p,q,y)$ is the Schur function for the single-strand supergraviton character
\begin{align} \label{eq.SGchar}
    z_{\mathrm{SG}}^{}(p,q,y) &= \sum_{k}p^k\Big[\phi^{(s)}_{\frac{k-1}{2}}(q,y) -2\phi^{(s)}_{\frac{k}{2}}(q,y) + \phi^{(s)}_{\frac{k+1}{2}}(q,y)\Big] \notag\\
    &\eqcolon \sum_{k,m,l} c_{\mathrm{SG}}^{}(k,m,l)\, p^kq^my^l\ ,
\end{align}
where $\phi^{(s)}_{j}$ are characters of short representations of $SU(1,1\,|\,2)$ (see \eqref{eq.su112short} of the Supplemental Material). 

In \cite{Maldacena:1999bp} the supergraviton spectrum was compared with the full CFT spectrum using the MEG \eqref{eq:MEG_T^4} and agreement was found below the black-hole threshold, $h<h_{\rm BH}=\frac{N}{4}$. However, this agreement is rather empty, as both the CFT and supergraviton MEGs in fact vanish below this threshold (except for the contribution of the global NS vacuum at order $q^0y^0$). We will now see that the REG \eqref{eq.REG1} turns this ``$0=0$'' statement into a meaningful comparison, by resolving each side of this equation into a sum of non-vanishing $\tj_2$-sector contributions \footnote{In version 1 of this Letter the proposed REG was based on sectors of contributions from Young diagrams $\lam\in H(1|1)$ with a fixed number of rows. While detailed matching with the supergraviton spectrum was also observed in that case, it is not clear how to prove the protection of that index. We have therefore defined the REG in the present version with respect to the weaker selection rules based on $\widetilde{su}(2)_2$ representations, whose protection at first order in the deformation is straightforward.}, $\cE_{\tj_2}$, with $\tj_2 = 0,
\frac12,\dots,\frac{N-1}{2}$. 
In order to make this comparison, the R-sector left-moving character of the CFT \eqref{eq:1-strand_z(p,q,y)} should be flowed to the NS sector via $z_{\rm NS}^{}(p,q,y)=z_{\rm  R}^{}(pq^{\frac12}y,q,yq^{\frac12})$~\cite{Schwimmer:1986mf}. 

At $N=3$, for example, the CFT and supergraviton REGs have the $q$-expansions \footnote{We acknowledge the use of the GAP system for computational discrete algebra \cite{GAP4}, which we used for the computation of symmetric group characters.}
\begin{align*} \label{eq:N=3REGs}
\begin{array}{r@{\,}l@{\,}c@{\,}l@{\,}c@{\,}l@{\,}c@{\,}l}
\mathcal{E}_{3,0}^{\mathrm{CFT}}     &= 3& + &q^{\frac12}(-6y-6y^{-1}) &+ &q\big(+4y^2+28+4y^{-2}) &+& \cdots \\
    \mathcal{E}_{3,\frac12}^{\mathrm{CFT}} &= &&q^{\frac12}(+6y+6y^{-1})& + &q(-12y^2-32 -12y^{-2}) &+& \cdots \\
    \mathcal{E}_{3,1}^{\mathrm{CFT}} &= &&&&q(+9y^2+12+9y^{-2}) &+& \cdots \\[1ex]
    \mathcal{E}_{3,0}^{\mathrm{SG}} &= 3\, &+ &q^{\frac12}(-6y-6y^{-1}) &+ &q(+4y^2+25+4y^{-2}) &+& \cdots \\
    \mathcal{E}_{3,\frac12}^{\mathrm{SG}} &= &&q^{\frac12}(+6y+6y^{-1})& + &q(-12y^2-36-12y^{-2}) &+& \cdots \\
    \mathcal{E}_{3,1}^{\mathrm{SG}} &= &&&&q(+9y^2+12+9y^{-2}) &+& \cdots
    \end{array}
\end{align*}
where we see not only detailed non-trivial matching up to the expected order $O(q^{\frac12})$, but also an enhanced matching for the $\tj_2=1$ sector, in which $\mathcal{E}_{3,\tj_2}^{\mathrm{CFT}}-\mathcal{E}_{3,\tj_2}^{\mathrm{SG}} = O(q^{\frac32})$. We have explicitly checked that these features (matching below the threshold and enhanced matching in certain $\tj_2$-sectors) of the REG hold up to $N=12$. This matching with the supergraviton spectrum is further evidence that lifted states cancel from the REG separately for each $\tj_2$-sector and therefore that the REG is protected.
\begin{figure}[b]
\begin{adjustbox}{center}
    \begin{subfigure}[h]{0.67\columnwidth}
        \includegraphics[height=4cm]{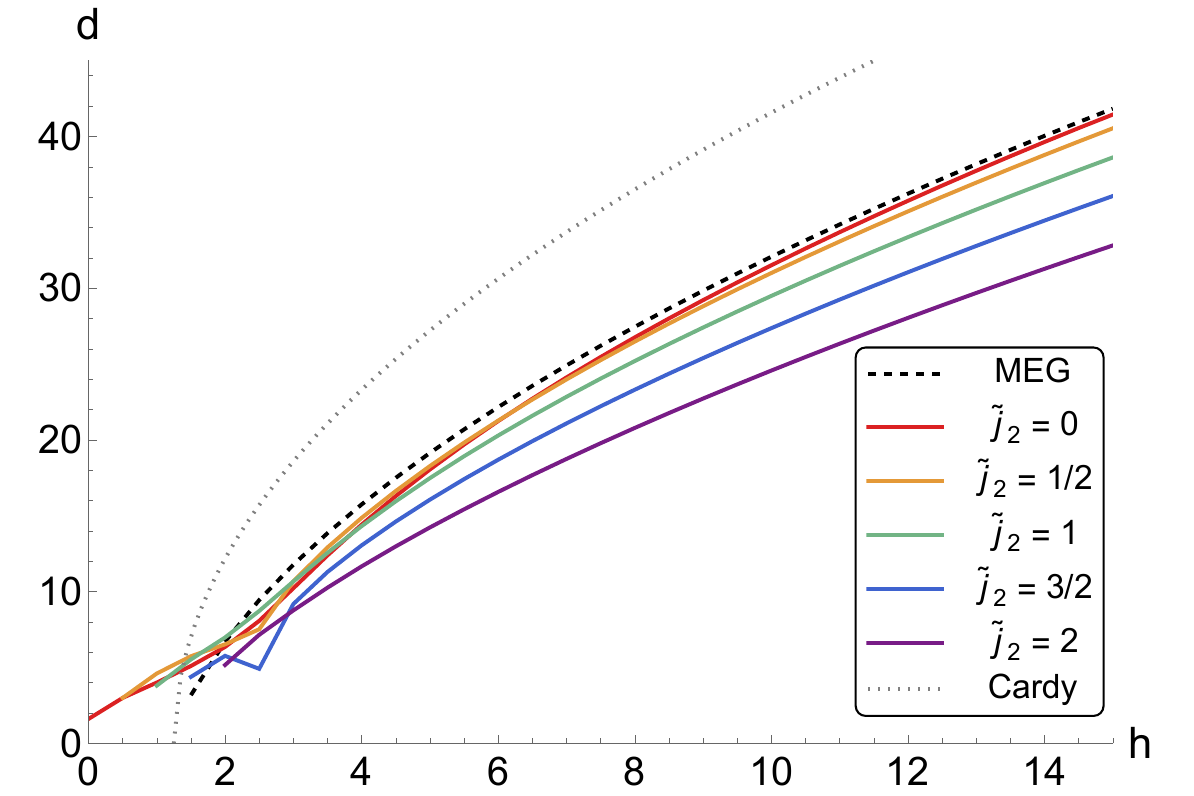}
        \caption{\label{fig:degensa}}
    \end{subfigure}
    \begin{subfigure}[h]{0.33\columnwidth}
        \includegraphics[height=4.15cm]{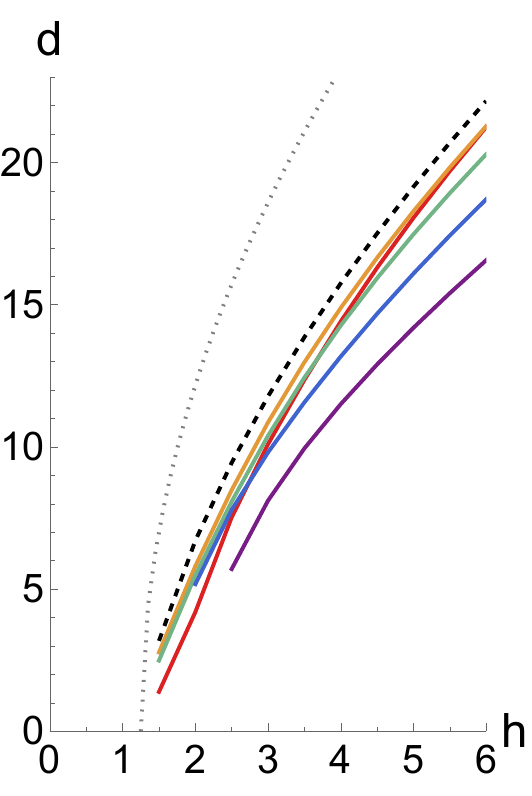}
        \caption{\label{fig:degensb}}
    \end{subfigure}
\end{adjustbox}
\caption{\raggedright\sl Plots of the REG logarithmic degeneracies $d^{\mathrm{CFT}}_{N=5,\tj_2}$ in (a) and $d^{\mathrm{BH}}_{N=5,\tj_2}$ in (b). For comparison, we show the analogous quantities obtained using the MEG, along with the universal Cardy growth. The first states contributing to the REG are at larger $h$ for sectors with larger $\tj$.\label{fig:degens}}
\end{figure}
In order to study states well above the black-hole threshold, we consider the logarithmic degeneracies of the CFT REG $d^{\mathrm{CFT}}_{N,\tj_2}(h) \coloneq \log\abs{\mathcal{E}_{N,\tj_2}^{\mathrm{CFT}}(q,1)|_{q^h}}$ as well as of the difference between the CFT and supergraviton REGs $d^{\mathrm{BH}}_{N,\tj_2}(h) \coloneq \log\abs{\mathcal{E}_{N,\tj_2}^{\mathrm{CFT}}(q,1)|_{q^h}-\mathcal{E}_{N,\tj_2}^{\mathrm{SG}}(q,1)|_{q^h}}$, where setting $y=1$ has the effect of summing over values of $J^3_0$. From the logarithmic degeneracies (shown for $N=5$ in Fig.~\ref{fig:degens}) we see that each $\tj$-sector of the REG appears to have the same leading-order growth of states as the full MEG in the regime $h\gg N$, namely Cardy growth \cite{Cardy:1986ie}. Since $\tj$-sectors contain a mixture of twist sectors, we conclude that black hole states are distributed among twist sectors. The subleading growths of the REG should, however, depend on $\tj$. It would be interesting to see what the more refined structure of the REG implies for the regime $h\sim N\gg1$ in which the ``long-string sector'' (the minimal $n_{\lambda}$ sector) is believed to be dominant~\cite{Maldacena:1996ds,Das:1996wn}. This long-string sector is contained within the $\tj=0$ sector of the REG. We also observe that for quantum numbers well into the black hole regime---that is, well within the parabola $h=\frac{j^2}{N}+\frac{N}{4}$ \cite{Breckenridge:1996is}---the coefficients of $q^hy^{2j}$ for fixed $h$ and $j$ appearing in the different $\tj$-sectors of the REG are all of the same sign.

\section{6.~~Outlook}

We constructed a new supersymmetry index for the D1-D5 CFT on $T^4$---the resolved elliptic genus (REG)---and argued for its protection using the precise form of the twisted-sector exactly marginal operators. We demonstrated detailed matching between the CFT and supergraviton spectra in each $\tj$-sector below the black-hole threshold---the first such matching since the MEG is trivial in this region~\cite{Maldacena:1999bp}. This is further evidence that the REG is protected. In fact, we found that many $\tj$-sectors even enjoy an enhanced CFT-supergraviton matching. Above the threshold, we showed that the REG resolves the black-hole microstates visible in the MEG into multiple distinct $\tj$-sectors.

The REG finds a natural description in the Schur-Weyl formalism developed in this Letter which, with minimal modifications, should be applicable to general symmetric orbifold theories---for studies spaces of such theories relevant to holography see \textit{i.e.}~\cite{Belin:2019rba}. In particular, it should be applicable to the D1-D5 CFT on $K3$; while the theory's Hilbert space (as defined from its torus orbifold description) does not totally factorize into left- and right-movers (contrary to the assumption of \eqref{eq:Hn_LxR}), a slightly modified formalism should still apply. We hope to report progress on this is the near future.

An interesting question is whether the REG has a gravity dual~\footnote{For recent developments on the bulk computations of supersymmetry indices, see {\it e.g.} \cite{Cabo-Bizet:2018ehj}}.  
The fugacity $x$ in \eqref{eq:REG_DMVV} counts representations of $\widetilde{su}(2)_2$ associated with the bulk $T^4$, but its bulk meaning is unclear and deserves further study.

Because of the D1-D5 CFT's central role in the study of black hole microstate physics, the REG opens a broad avenue of applications and investigations, potentially providing a new handle on longstanding and actively pursued problems---including the lifting problem \cite{Hampton:2018ygz,Benjamin:2021zkn,Guo:2019pzk,Gaberdiel:2024nge,Guo:2022ifr}, fortuity \cite{Chang:2025rqy}, BPS chaos \cite{Chen:2024oqv}, the holographic dictionary for multi-center configurations \cite{Dabholkar:2009dq, Bena:2011zw, Denef:2007vg},
AdS$_3\times S^3\times S^3\times S^1$ holography \cite{Murthy:2025moj,Eberhardt:2017pty,Eberhardt:2018sce},
and the relation to the generalized supergravity index \cite{Hughes:2025car}. We hope to report progress on these subjects in the future.

\vspace*{2ex}
\noindent\textbf{Acknowledgments.}
We would like to thank Hiroaki Kanno for fruitful discussion. We thank Rodolfo Russo, Stefano Giusto, David Turton, Samir Mathur and Bin Guo for helpful comments on version 1 of this paper.
We thank CEA Saclay where this work was partially done during the workshop``Black-Hole Microstructure VII.''
We also thank Yukawa Institute for Theoretical Physics at Kyoto University, where this work was partially done during the workshop YITP-I-25-01 on ``Black Hole, Quantum Chaos and Quantum Information.''
This work was supported in part by MEXT
KAKENHI Grant Numbers 21H05184 and 24K00626.


\bibliography{NewIndexPRD}

\onecolumngrid

\clearpage
\appendix*
\section*{Supplemental Material}

The character of short representations of global $SU(1,1\,|\,2)$ is given by
\begin{equation} \label{eq.su112short}
    \phi^{(s)}_{j}(q,y) = \frac{q^j}{1-q}\frac{y^{2j}(y-2\sqrt{q}+y^{-1}q)-y^{-2j}(y^{-1}-2\sqrt{q}+yq)}{y-y^{-1}}\ ,\qquad  
    j=0,\frac12,1,\dots\ .
\end{equation}

The theta function and the Dedekind eta function are defined as
\begin{align} \label{eq.theta1etaDef}
 \vartheta_{1}(\nu,\tau)
 \coloneq-iq^{\frac{1}{8}}\big(y^{\frac12}-y^{-\frac12}\big) \prod_{m=1}^{\infty} (1-q^m)(1-zq^{m})(1-z^{-1}q^{m})\ ,\qquad\ 
 \eta(\tau)\coloneq q^{\frac{1}{24}}\prod_{m=1}^{\infty}(1-q^m)\ ,
\end{align}
where $q=e^{2\pi i\tau}$, $y=e^{2\pi i \nu}$.

\bigskip
\noindent{\bf Schur functions and power sum polynomials}\\
The (bosonic) Schur polynomials $S_\lambda(x)$ give a basis for homogeneous symmetric polynomials of degree $n$ in $b$ variables $x_1,\dots,x_b$, and are indexed by a partition $\lambda$ of $n$ at most into $b$ parts: \textit{i.e.} 
$\lambda=(\lambda_1\ge \lambda_2\ge \cdots \ge \lambda_b \ge 0)$, $\lambda_1+\cdots+\lambda_b=n$, which can be represented by a Young diagram with $n$ boxes and up to $b$ rows.
It is usual to take $b$ to be large, at least $b\ge n$, or often $b=\infty$, in which case Schur polynomials are called Schur functions.

The power sum polynomials give another basis of symmetric functions and are defined by
\begin{align}
 P^{({\bf i})}(x) \coloneq P_1(x)^{i_1}\,\cdots\, P_n(x)^{i_n}\ ,
\qquad P_j(x) \coloneq x_1^j+\cdots+x_b^j\ ,
\label{eq:def_power_sum}
\end{align}
where ${\bf i}=(i_1,\dots,i_n)$, $\sum_\alpha \alpha i_\alpha = n$, is a partition of $n$.
Schur functions can be expanded in terms of $P^{({\bf i})}(x)$ as
\begin{equation}
 S_\lambda(x) = \sum_{{\bf i}\,\vdash n}\frac{\omega_\lambda({\bf i})}{z({\bf i})}P^{({\bf i})}(x)\ ,\qquad
 z({\bf i}) \coloneq
 \prod_{\alpha=1}^n i_\alpha!\,\alpha^{i_\alpha}\ ,
\label{Schur_ito_power_sum}
\end{equation}
where 
\begin{equation}
\omega_\lambda({\bf i})
=\big[\Delta(x)\, P^{({\bf i})}(x)\big]_l\ ,
\qquad \Delta(x)=\prod_{i<j}(x_i-x_j)\ ,\qquad
l\coloneq (\lambda_1+b-1,\lambda_2+b-2,\dots,\lambda_b)\ ,
\label{eq:def_omega_lambda}
\end{equation}
and
$[Q(x)]_\lambda \coloneq (\text{coefficient of $x_1^{\lambda_1}\cdots\, x_n^{\lambda_n}$})$
for a symmetric polynomial $Q(x)$.

Physically, $P_1(x)$ can be regarded as the trace of some operator $g$ with eigenvalues $x_1,\dots,x_b$ in an $n$-dimensional Hilbert space; namely, $P_1(x)=\tr[g]$.  More generally, $P_j(x)={\tr}[g^j]$.

A super Schur function \cite{Macdonald1992} (or hook Schur function) $S_\lambda(x|\tilde{x})$ is a function of two sets of variables $x_1,\dots,x_b$ and $\tilde{x}_1,\dots,\tilde{x}_f$ and, for our purposes, we can define them to be a function given by
\eqref{Schur_ito_power_sum} with $P^{({\bf i})}(x)$ replaced by the super version
\begin{align}
 P^{({\bf i})}(x|\tilde{x}) \coloneq P_1(x|\tilde{x})^{i_1}\,\cdots\, P_n(x|\tilde{x})^{i_n}\ ,
\qquad 
P_j(x|\tilde{x})\coloneq(x_1^j+\cdots+ x_b^j)
-(\tilde{x}_1^j+\cdots+\tilde{x}_f^j)\ ,
\label{eq:def_super_power_sum}
\end{align}
with $\omega_\lambda({\bf i})$ still given by the formula \eqref{eq:def_omega_lambda} in terms of the bosonic $P^{({\bf i})}$.

Physically, if $\cH$ is a Hilbert space of $b$ bosonic and $f$ fermionic dimensions, $P_1(x|\tilde{x})$ can be regarded as the trace with $(-1)^F$ in $\cH$ of some operator $g$ with eigenvalues $x_1,\dots,x_b$ in the bosonic subspace and $\tilde{x}_1,\dots,\tilde{x}_f$ in the fermionic subspace; namely, $P_1(x|\tilde{x})={\tr}[(-1)^F g]$.  More generally, $P_j(x|\tilde{x})={\tr}[(-1)^Fg^j]$.   If $g=p^{\hat{k}}q^{L_0}y^{2J^3_0}$ as below \eqref{eq:1-strand_char_L}, and if we write $P_1={\tr}[(-1)^Fg]=z_1(p,q,y)$, then $P_j={\tr}[(-1)^F g^j]=z_1(p^j,q^j,y^j)$.  This is used to evaluate \eqref{eq.REG1}.

The super Schur functions satisfy
the Cauchy identity
\begin{equation}
    \sum_\lambda S_\lambda(x|\tilde x)S_\lambda(x'|\tilde x')
    =
    \frac{\prod_{i,\tilde{\imath}'}(1-x_i \tilde{x}'_{\tilde{\imath}'})
    \prod_{i',\tilde{\imath}}(1-x'_{i'}\tilde{x}_{\tilde{\imath}})
    }{
    \prod_{i,\tilde{\imath}}(1-x_i \tilde{x}_{\tilde{\imath}})
    \prod_{i',\tilde{\imath}'}(1-x'_{i'}\tilde{x}'_{\tilde{\imath}'})
    }\ .
    \label{eq:Cauchy_id_general_app}
\end{equation}
where the sum is over all Young diagrams $\lambda$ with any number of boxes.

\end{document}